\documentclass[11pt]{article}
\usepackage{graphicx}
\usepackage{longtable,lscape}
\usepackage{amsfonts}
\usepackage{bbm}
\usepackage{amsmath,amsthm}

\usepackage{float}
\usepackage{url}

\newcommand{\captionfonts}{\footnotesize}
\makeatletter  
\long\def\@makecaption#1#2{%
  \vskip\abovecaptionskip
  \sbox\@tempboxa{{\captionfonts #1: #2}}%
  \ifdim \wd\@tempboxa >\hsize
    {\captionfonts #1: #2\par}
  \else
    \hbox to\hsize{\hfil\box\@tempboxa\hfil}%
  \fi
  \vskip\belowcaptionskip}
\makeatother 
\begin{document}
\title{A Generalized Probability Framework to Model Economic Agents' Decisions Under Uncertainty}
\author{Emmanuel Haven \ and \ Sandro Sozzo \\
         School of Management and IQSCS\\
				 University of Leicester \\
        University Road LE1 7RH Leicester, UK \\
        E-Mails: \url{eh76,ss831@le.ac.uk}} 

\maketitle
\begin{abstract}
The applications of techniques from statistical (and classical) mechanics to
model interesting problems in economics and finance has produced valuable
results. The principal movement which has steered this research direction is
known under the name of `econophysics'. In this paper, we illustrate and
advance some of the findings that have been obtained by applying the
mathematical formalism of quantum mechanics to model human decision making
under `uncertainty' in behavioral economics and finance. Starting from
Ellsberg's seminal article, decision making situations have been
experimentally verified where the application of Kolmogorovian probability
in the formulation of expected utility is problematic. Those probability
measures which by necessity must situate themselves in Hilbert space (such
as `quantum probability') enable a faithful representation of experimental
data. We thus provide an explanation for the effectiveness of the
mathematical framework of quantum mechanics in the modeling of human
decision making. We want to be explicit though that we are not claiming that
decision making has microscopic quantum mechanical features.
\end{abstract}

\section{Introduction\label{intro}}

Roughly speaking, `econophysics' concerns the application of classical (and statistical mechanical) physics theories to model the behavior of economic and financial systems. The econophysics movement has been leaded by several brilliant physicists (see, e.g., \cite{1,2,3,4,5}). This article aims to bring to the attention of
econophysicists a novel emerging domain where the application of methods and
techniques inspired by quantum physics has been successful in the last
years. This domain, known in the scientific community as `quantum cognition',
was born as a bold proposal to solve a specific problem.\footnote{%
See, e.g,. the articles \textquotedblleft Quantum minds: Why we think like
quarks\textquotedblright , \textit{New Scientist} 05 September 2011, by M.
Buchanan, and \textquotedblleft Physics goes social: How behavior obeys
quantum logic\textquotedblright , \textit{New Scientist} 11 July 2013, by A.
Khrennikov and E. Haven.}

The quantum cognition domain applies the mathematical formalism of quantum
mechanics to model situations and processes in human cognition, decision
making and language that have resisted traditional modeling techniques by
means of classical structures, i.e. Boolean logical structures,
Kolmogorovian probability spaces, Bayesian update of probabilities, commutative algebras, etc.
(please see Sect. \ref{quantumcognition}). Therefore, the results
obtained in quantum cognition have a deep impact on behavioral economics
and finance. This domain has attracted in the last years the interest of
high impact factor and top journals, media and popular science and
funding institutions \cite{azz2009,a2009a,k2010,bpft2011,bb2012,abgs2013,ags2013,hk2013,ast2014,ys2014,s2014a,s2014b}. To better clarify the boundaries of quantum cognition it is worth mentioning two important aspects of it, which are as follows.

(i) The success of this quantum modeling is interpreted as due to the
`descriptive effectiveness of the mathematical apparatus of quantum theory
as a formal tool to model cognitive dynamics and structures in
situations where classical set-based approaches are problematical',
`without any' a priori direct or precise connection with the validity of
quantum laws in the microscopic world.

(ii) There is no need, in order to guarantee the validity of the obtained
results, to introduce any compelling assumption about the existence of
microscopic quantum processes at the level of the human brain. Hence,
quantum cognition should not be confused with `quantum mind' or `quantum
consciousness'.

What are the possible advantages of quantum cognition in economics? In this
respect, the application of normative models of decision making to the
behavior of economic agents has produced a variety of sophisticated
mathematical frameworks, the most important of which are `expected utility
theory' (EUT) \cite{vonneumannmorgenstern1944} and `subjective expected
utility theory' (SEUT) \cite{savage1954}. The former is designed for
decisions under `risk', that is, a choice among different gambles defined on an
objective probability measure, whilst the latter is designed for decisions
under `ambiguity', that is, a choice among different acts defined on a
subjective probability measure. Both theories implicitly assume that
`probabilities are Kolmogorovian', that is, probabilities are assigned to
events according to rules obeying the axioms of Kolmogorov. However, since
the work of Maurice Allais \cite{allais1953}, decision economists
systematically produce empirical situations where concrete human decisions
violate the axioms of EUT. Moreover, since the work of Daniel Ellsberg \cite%
{ellsberg1961}, decision economists are also able to generate empirical
situations where concrete human decisions violate the axioms of SEUT.
Finally, recent work of Mark Machina \cite{machina2009} reveals that the
most recognised extensions of EUT and SEUT able to cope with `Allais' or
`Ellsberg paradoxes' are highly problematical in specific decision making
situations, i.e. `Machina paradox' (please see Sect. \ref{ellsberg}).

Inspired by our quantum cognition approach, we have recently elaborated a complete modeling of the `Ellsberg
paradox' by using the mathematical formalism of quantum mechanics \cite{ast2014,hk2009,ast12}. We have also faithfully represented the data collected in
an experiment we performed on a typical Ellsberg paradox situation with real
decision makers \cite{ast2014}. In this paper we further inquire into our quantum-theoretic framework for the Ellsberg paradox, showing that our results go beyond the mere theoretical modeling and representation of a set of empirical data. We also provide sufficient arguments to claim that, not only in the Ellsberg paradox, but also in other situations affected by ambiguity, such as the `Machina paradox', structurally there is a real need for a non-classical probability model. We would like to advance two reasons.

(i) In an Ellsberg-type decision making process, the agent's choice is actualized as a consequence of an interaction with the cognitive context, exactly like in a quantum measurement process where the outcome of the measurement is actualized as a consequence of the interaction of the measured particle with the measuring apparatus. Therefore, in cognitive entities, as well as in microscopic quantum entities, measurements do not reveal preexisting values of the observed properties but, rather, they actualize
genuine potentialities. Classical Kolmogorovian probability can only formalize lack of knowledge about actualities, hence it is generally not able to cope with a decision making process.  We have proven that this is possible by using a complex Hilbert space, and by
representing probability measures by means of `projection valued measures'
on a complex Hilbert space \cite{a2009a,s2014a,s2014b}. A projection valued measure is essentially
different from a single Kolmogorovian probability measure, since the latter
is a $\sigma $--algebra valued measure, whilst the former is not. 

(ii) The notion of ambiguity, as introduced in economics, is completely compatible, both at a mathematical and an intuitive level, with the representation of states of cognitive entities as vectors of a Hilbert space. Indeed, just like in standard quantum mechanics the state vector incorporates the `quantum uncertainty' of a microscopic particle, also in an Ellsberg-type situation, the agent's subjective preference towards ambiguity is naturally formalized by representing the state of the cognitive entity under study by means
of such a Hilbert space vector (this perspective is getting more and more accepted in the scientific community, including mainstream psychologists; please see \cite{pnas}). In this respect, it is worth mentioning that Ellsberg called `ambiguity aversion' the `irrational' factor inducing decision makers to deviate from SEUT. In our approach, ambiguity aversion is only one -- albeit an important one -- of the conceptual landscapes surronding the decision maker's choice in a situation where ambiguity is present. This result is compatible with the experimental findings confirming Ellsberg's prediction about the human attitude towards ambiguity \cite{ellsbergexp1}, but also with some recent experiments where such attitude is more controversial \cite{ellsbergexp2}.

Points (i) and (ii) provide an intuitive explanation for the identification of genuine quantum structures in the Ellsberg paradox. Those structures are typically characterized by notions such as `contextuality', `interference' and `superposition', which will be discussed in more detail in Sect. \ref{quantumellsberg}. 

In concluding this section, it is important to mention that our model which aims to represent human decision making in economics is a descriptive model: it describes what economic agents actually do, not what they should do, under uncertainty. However, it already contains some insights on how the construction of an axiomatic framework of what we could call `contextual expected utility' as based on a non-classical probability can be able to cope with human ambiguity, or `contextual risk', as we could call it. If we wanted to embed our approach into the fundamentals of microeconomics, then a natural generalization of EUT and SEUT may simply consist in requiring that economic agents maximize their contextual expected utility. An important achievement in that regard would require a representation theorem which provides for a rigorous proof of the equivalence between the existence of a preference relationship and an order inequality between utility functions embedding this type of expected utility.

Our generalization of the probability models 
employed in an expected utility framework has a profound impact on any economics or finance problem where this basic microeconomic framework is used as an input in its modeling objectives. Indeed, an important assumption in general equilibrium based macroeconomic models is the `rational expectations hypothesis' which exactly rests on the expected utility hypothesis. The consistency of the models imposed by rational expectations has profound implications on the design and impact of macroeconomic policy-making \cite{HansenSargent2000,equity}.

\section{On the effectiveness of quantum mathematics in human cognition\label{quantumcognition}}
Classical Boolean logic and Kolmogorovian (or Bayesian) probability theory have exercised a long influence on the way in which scholars formalize human behavior under uncertainty. However, empirical evidence, accumulated in the last thirty years in cognitive psychology, clearly indicates that these classical structures do not provide the most general modeling framework for human decision making.

There are three major domains of cognition where deviations from classical logical and probabilistic structures have been observed.

The first of these two domains is `concept theory'. Cognitive scientists know that
concepts exhibit `graded', or `fuzzy', `typicality', e.g., humans estimate
an exemplar such as \textit{Robin} as more typical than \textit{Stork} as a
typical example of the concept \textit{Bird}. A problem arises when one
tries to formalize the typicality of the combination of two concepts in
terms of the typicality of the component concepts which form that
combination. One is intuitively led to think that the standard rules of classical
(fuzzy set) logic and probability theory apply in such combinations. However, Osherson and Smith showed in
1981 that this intuition is not correct for concept conjunctions. Humans
score the typicality of an exemplar such as \textit{Guppy} with respect to
the conjunction \textit{Pet-Fish} as higher than the typicality of \textit{%
Guppy} with respect to both \textit{Pet} and \textit{Fish} separately
(`Guppy effect') \cite{os1981}. One realizes at once that typicality
violates standard rules of classical (fuzzy set) logic. A
second set of human experiments on concept combinations was performed by
James Hampton. He measured the membership weight, i.e. normalized membership
estimation, of several exemplars, e.g., \textit{Apple}, \textit{Broccoli}, 
\textit{Almond}, etc., with respect to pairs of concepts, e.g., \textit{%
Fruits}, \textit{Vegetables}, and their conjunction, e.g., \textit{Fruits
and Vegetables}, or disjunction, e.g., \textit{Fruits or Vegetables}. These
membership weights again showed systematic deviations from standard classical (fuzzy
set) rules for conjuction \cite{h1988a} and disjunction of two concepts \cite%
{h1988b}. The conclusion is immediate: one
cannot express conceptual gradeness in a classical (fuzzy) set-theoretic
model. In more general terms, one cannot represent experimental membership
weights in a single classical probability space satisfying the axioms of
Kolmogorov \cite{a2009a}.

The second set of empirical findings showing unexpected deviations from
classicality pertains to `decision theory' and can be traced back to the
work of Kahneman and Tversky in the 1980's. Their famous experiment on the
`Linda story' revealed that situations exist where the participants estimate
the probability of the conjunction of two events as higher than the
probability of one of them, thus violating monotonicity of classical
probability (more generally, Bayes' rule) \cite{tk1983}. This `conjunction
fallacy' is an example of a `human probability judgement' (a `disjunction
fallacy' has also been observed where humans estimate the probability of the
disjunction of two events to be less than the probability of one of them)
and classical Kolmogorovian probability is again not appliable in these
cases \cite{bb2012}. A `disjunction effect' was also identified by Tversky
and Shafir in the nineties \cite{ts1992}. In the latter effect, people
prefer action $A$ over action $B$ if they know that an event $X$ occurs, and
also if they know that $X$ does not occur, but they prefer $B$ over $A$ if
they do not know whether $X$ occurs or not. The disjunction effect violates
a fundamental principle of expected utility theory, Savage's `Sure-Thing
principle' \cite{savage1954} (more generally, the total probability law of
classical probability), revealing that humans show an `uncertainty
aversion', as we will see in Sect. \ref{ellsberg}.

These surprising findings led several scholars to explore alternatives to
traditional modeling approaches that could better cope with the effects,
fallacies and paradoxes above. As we already have tried to argue in Sect. \ref{intro}, a major alternative can be the so called
`quantum cognition approach', which applies the conceptual and mathematical
framework of quantum mechanics to model cognitive processes \cite%
{azz2009,a2009a,k2010,bpft2011,bb2012,abgs2013,ags2013,hk2013,ast2014,s2014a,ys2014,s2014b}%
.

Quantum mechanics provides a specific mathematical framework to formalize
microscopic phenomena. More concretely, any entity is described in the quantum formalism by a specifically structured
linear space over complex numbers, called `Hilbert space'. The state of an
entity is represented by a vector belonging to this Hilbert space, while a
measurement performed on an entity by means of a measurement apparatus is
represented by a specific operator, called `self-adjoint operator', mapping
a vector of this Hilbert space into another vector of the same space. The
measuring apparatus provides a `measurement context' for the measured entity
and induces an `indeterministic change of state' of the entity itself, in
which a single outcome is actualized in a set of possible outcomes as a
consequence of the interaction between the measured entity and the measuring
apparatus. Heisenberg called this change of state a `transition from
potential to actual', since the quantum state incorporates these intrinsic
and unavoidable aspects of `contextuality', `pure potentiality' and
`uncertainty'. The statistics of repeated measurements is described by a
probabilistic rule, called the `Born rule', and the ensuing quantum
probability model does not satisfy the restrictions of classical
Kolmogorovian probability. This mathematical formalism naturally copes with
very fundamental quantum effects, such as `interference', `superposition' and `entanglement'.

What was a priori completely unexpected is that the mathematical formalism
sketched above, which we call `quantum mathematics' to emphasize that it is
used for modeling purposes outside of physics, has been very powerful
to represent human decision making. In particular, the theoretical framework
of quantum probability, which is more general than classical Kolmogorovian
probability, has been able, not only to cope with all the cognitive
situations above, where traditional probability models are problematical,
but it has also shown a capability to predict the results of new experiments
which have been performed in recent years. In our opinion, those outcomes do
show that there is a serious rationale for arguing that genuine quantum
structures may exist in the mechanisms of conceptual combination \cite%
{a2009a,abgs2013,ags2013,s2014a,s2014b}, human probability and similarity
judgments \cite{azz2009,bpft2011,bb2012,ys2014,pb2009}, and thereby also in
behavioral economics/finance \cite%
{k2010,hk2013,ast2014,pb2009,danilovlambert2010}.  

The success of the quantum cognition
approach goes beyond its modeling effectiveness. Indeed, there are deep
analogies between the interactions (of a physical nature) occurring in a
quantum measurement process, and the interactions (of a cognitive nature)
occurring in a decision process. More concretely, in a quantum measurement process, the measurement
context actualizes one outcome among the possible outcomes, thus provoking
an indeterministic change of state of the microscopic quantum particle that
is measured. Similarly, whenever a person is asked to give a preference, or
to make a choice, or to take a decision, contextual influence (of a
cognitive type) and a transition from potential to actual occur in which an
outcome is actualized from a set of possible outcomes. One can say that, in
both quantum and cognitive realms, measurements `create', rather than just `record', properties of the measured
entities. A second important common aspect of quantum and cognitive realms
is that each measurement changes the state of the measured entity in a
different way, that is, the probability of getting a given pair of outcomes
in two sequential measurements depends on the order in which the
measurements are performed. One typically says that measurements are
generally `non-commutative', or `non-compatible', in the quantum jargon.
Cognitive experiments frequently provide significant examples of
non-compatible sequential questions. At variance with classical
Kolmogorovian probability, quantum probability enables coping with this kind
of contextuality, pure potentiality and order effects occurring in both
physical and cognitive realms \cite{a2009a,bb2012,ags2013}.

The result is that we are at a theoretical crossroad in human cognition,
having either to continue relying on classical probability theory and accept
the observed deviations as fallacies or aversions, or to look at alternative
approaches that allow for the provision of better descriptive and normative
accounts of human decision making. The chosen path may have an impact on human life in general, as economics, financial and political  decisions strongly depend on decision making models.

In this framework we can situate the third set of empirical difficulties of traditional classical probabilistic approaches to human decision making, namely, those characterizing behavioral economics. This will be discussed in the next section.

\section{The Ellsberg paradox\label{ellsberg}}

It is usually maintained that individuals behave in uncertainty situations
in such a way that they maximize their wealth which, according to `expected
utility theory' (EUT) \cite{vonneumannmorgenstern1944,savage1954}, can be
achieved by maximizing expected utility. The simplicity, mathematical
tractability and predictive success of EUT make it the predominant economics model of decision making under uncertainty. However,
starting from the work of Maurice Allais \cite{allais1953}, significant
empirical deviations from this `expected utility maximization rule' have been systematically
observed in specific types of situations, and these deviations are
traditionally referred to as paradoxes.

EUT was originally developed by von Neumann and Morgenstern \cite%
{vonneumannmorgenstern1944}. They formulated a set of axioms that allow to
represent decision maker preferences over the set of `acts' (functions from
the set of states of nature into the set of consequences) by a suitable
functional $E_{p}u(.)$, the `expected utility', where $u$ is a `Bernouilli
utility function' on the set of consequences, and $p$ is an objective
probability measure on the set of states of nature.

How does EUT cope with uncertainty? Knight pointed out the difference
between `risk' and `uncertainty' reserving the term `risk' for situations
that can be described by objective probabilities, and the term `uncertainty' to refer
to situations in which agents do not know the probabilities associated with
each of the possible outcomes of an act \cite{knight1921}. Von Neumann and
Morgenstern's formulation of EUT did not contemplate the latter possibility,
since all probabilities are objective in their
scheme. For this reason, Savage extended EUT allowing agents to construct
their own subjective probabilities when objective probabilities are not
available \cite{savage1954}. Hence, according to Savage's extension of EUT,
or SEUT, the distinction proposed by Knight would be irrelevant. Ellsberg
instead showed in a series of `thought experiments' that Knightian's
distinction is empirically meaningful, thus pointing out some limitations of
Savage's SEUT \cite{ellsberg1961}. In particular, Ellsberg presented the
following experiment.

Consider one urn with 30 red balls and 60 balls that are either yellow or
black, the latter in unknown proportion. One ball will be drawn at random
from the urn. Then, free of charge, a person is asked to bet on one of the
acts $f_1$, $f_2$, $f_3$ and $f_4$ defined in Tab. 1. \noindent 
\begin{table}[tbp]
\label{table01}
\par
\begin{center}
\begin{tabular}{|p{1.5cm}|p{1.5cm}|p{1.5cm}|p{1.5cm}|}
\hline
Act & red & yellow & black \\ \hline\hline
$f_1$ & \$12 & \$0 & \$0 \\ \hline
$f_2$ & \$0 & \$0 & \$12 \\ \hline
$f_3$ & \$12 & \$12 & \$0 \\ \hline
$f_4$ & \$0 & \$12 & \$12 \\ \hline
\end{tabular}%
\end{center}
\caption{The payoff matrix for the typical Ellsberg paradox situation.}
\end{table}
\noindent When asked to rank these gambles most of the persons choose to bet
on $f_1$ over $f_2$ and $f_4$ over $f_3$. This `Ellsberg preference' cannot
be explained by SEUT. Indeed, preferences must be consistent under SEUT, in
the sense that $f_1$ is preferred to $f_2$ if and only if $f_3$ is preferred
to $f_4$. Rephrasing, individuals' ranking of the sub-acts [12 on `red'; 0
on `black'] versus [0 on `red'; 12 on `black'] depends upon whether the
event `yellow' yields a payoff of 0 or 12, contrary to what is suggested by
the Sure-Thing principle, one of the axioms of Savage's SEUT.\footnote{%
The Sure-Thing principle was presented by Savage by introducing the
`businessman example' and can be rigorously formalized within SEUT, but it
can be intuitively stated as in Sect. \ref{quantumcognition}.} The
conclusion follows at once. There is no way to define an utility function $u$%
, associated with the given payoffs, and subjective probabilities,
associated with the events `red', `yellow' and `black', such that the
preferences observed in the Ellsberg situation are satisfied. Nevertheless,
these choices have a direct intuition: $f_1$ offers the 12 prize with an
`objective probability' of $1/3$, and $f_2$ offers the same prize but in an
element of the `subjective partition' $\{$`black', `yellow'$\}$. In the same
way, $f_4$ offers the prize with an objective probability of $2/3$, whereas $%
f_3$ offers the same payoff on the union of the unambiguous event `red' and
the ambiguous event `yellow'. Thus, in both cases the unambiguous bet is
preferred to its ambiguous counterpart. This preference for known
probability over ambiguous bets is now called `ambiguity aversion'.
Interestingly enough, the deviation from classical logical reasoning
observed in the Ellsberg paradox is deeply connected with the disjunction
effect introduced in Sect. \ref{quantumcognition}, both being charactedrized
by a violation of the Sure-Thing principle.

Extensions of SEUT were worked out to tackle the issues of SEUT raised by
Ellsberg-type preferences. These generalizations are primarily axiomatically
formulated and consist in replacing the Sure-Thing principle by weaker
axioms. We briefly summarize the most known extensions of SEUT, as follows.

(i) `Choquet expected utility'. This model considers a subjective
non-additive probability (or, capacity) over the states of nature rather
than a subjective probability. Thus, decision makers could underestimate or
overestimate probabilities in the Ellsberg experiment, and ambiguity
aversion is equivalent to the convexity of the capacity (pessimistic
beliefs) \cite{gilboa,schmeidler,choquet}.

(ii) `Max-Min expected utility', or `expected utility with multiple priors'.
The lack of knowledge about the states of nature of the decision maker
cannot be represented by a unique probability measure but, rather, by a set
of probability measures. Then, an act $f$ is preferred to $g$ if and only if 
$\min_{p\in P} E_p u(f) > \min_{p\in P} E_p u(g)$, where $P$ is a convex and
closed set of additive probability measures. Ambiguity aversion is
represented by the pessimistic beliefs of the agent which takes decisions
considering the worst probabilistic scenario \cite{gilboaschmeidler1989}.

(iii) `Variational preferences'. In this dynamic generalization of the
Max-Min expected utility, agents rank acts according to the criterion $%
\inf_{p\in \bigtriangleup } \{E_p u(f)+c(p)\}$, where $c(p)$ is a closed and
convex penalty function associated with the probability election \cite%
{mmr2006}.

(iv) `Second order probabilities'. This is a model of preferences over acts
where the decision maker prefers act $f$ to act $g$ iff $E_{\mu} \phi (E_p u
(f) ) > E_{\mu} \phi$ $(E_p u (g))$, where $E$ is the expectation operator, $%
u$ is a von Neumann-Morgenstern utility function, $\phi$ is an increasing
transformation, and $\mu$ is a subjective probability over the set of
probability measures $p$ that the decision maker thinks are feasible.
Ambiguity aversion is here represented by the concavity of the
transformation $\phi$ \cite{kmm2005}.

The approaches (i)--(iv) have been extensively applied in economic and
financial modeling. Given the enormous challenge of formalizing human
decision making, it does not come as a surprise that the four above
approaches do have shortcomings (\cite{machina2009,epstein1999}). It should
also be stressed that in fact none of these models can satisfactorily
represent more general Ellsberg-type situations, e.g., `Machina-type
paradoxes' \cite{machina2009,bdhp2011}). 

In 2009 Mark Machina proposed new thought experiments, the `50:51 example' and the `reflection example', which seriously challenged the approaches (i)--(iv) \cite{machina2009,bdhp2011}). In particular, the reflection example questions the `tail separability' assumption of Choquet expected utility exactly as the Ellsberg three-color example questions the Sure-Thing principle of SEUT. A version of the Machina reflection example can be formalized as follows.

Consider one urn with 20 balls, 10 are either red or yellow in unknown proportion, 10 are either black or green in unknown proportion. One ball will be drawn at random from the urn. Then, free of charge, a person is asked to bet on one of the acts $f_1$, $f_2$, $f_3$ and $f_4$ defined in Tab. 2. 
\noindent 
\begin{table} \label{table02}
\begin{center}
\begin{tabular}{|p{1.5cm}|p{1.5cm}|p{1.5cm}|p{1.5cm}|p{1.5cm}|}
\hline
Act & Red & Yellow & Black & Green \\ 
\hline
\hline
$f_1$ & \$0 & \$50 & \$25 & \$25 \\ 
\hline
$f_2$ & \$0 & \$25 & \$50 & \$25 \\ 
\hline
$f_3$ & \$25 & \$50 & \$25 & \$0 \\ 
\hline
$f_4$ & \$25 & \$25 & \$50 & \$0 \\ 
\hline
\end{tabular}
\end{center}
\caption{The payoff matrix for the `Machina reflection example with lower tail shifts'.}
\end{table}
\noindent 
Machina introduced the notion of `informational symmetry', that is, the events ``the drawn ball is red or yellow'' and ``the drawn ball is black or green'' have known and equal probability and, furthermore, the ambiguity about the distribution of
colors is similar in the two events. Under informational symmetry, people should prefer act $f_1$ over act $f_2$ and act $f_4$ over act $f_3$, or they should prefer act $f_2$ over act $f_1$ and act $f_3$ over act $f_4$. This is in particular inconsistent with the predictions of Choquet expected utility in (i). A recent experiment confirms the preference of $f_1$ over $f_2$ and  $f_4$ over $f_3$, consistently with informational symmetry \cite{machinaexp}.

We do not insist on the Machina paradox in the present paper, for the sake of brevity -- we will briefly come back to it in Sect. \ref{quantumellsberg} within our quantum-theoretic approach. We limit ourselves to mention that further investigation is still needed towards the construction of a satisfactory framework for representing ambiguity, ambiguity aversion
and, more generally, human preferences under uncertainty \cite{machinaexp}.

\section{Quantum structures in the Ellsberg paradox\label{quantumellsberg}}

We expose in this section our approach to economic agents decision making
based on the mathematical formalism of quantum mechanics. We do not present
the technical details of our modeling, but instead we aim to be as intuitive
and explicative as possible.The reader interested to the technical aspects
of our approach can refer to our papers \cite{ast2014,hk2009,ast12}. 

The first insight towards the elaboration of a quantum probabilistic
framework to model Ellsberg-type situations came from our conceptual and
structural investigation of how the approaches generalizing SEUT cope with
ambiguity and ambiguity aversion \cite%
{gilboa,gilboaschmeidler1989,mmr2006,kmm2005}. As we know, ambiguity
characterizes a situation without a probability model describing it, while
risk characterizes a situation where one presupposes that a classical
probability model on a $\sigma$--algebra of events exists. The
generalizations in (i)--(iv), Sect. \ref{ellsberg}, consider more general
structures than a single classical probability model on a $\sigma$--algebra.
We are convinced that this is exactly the point: ambiguity, due to its
contextual nature, structurally needs a non-classical probability model. To
this end we have elaborated a general framework for this type of situations,
based on the notion of `contextual risk' and inspired by the probability
structure of quantum mechanics. The latter is indeed intrinsically different
from a classical probability on a $\sigma$--algebra, because the set of
events does `not' form a Boolean algebra (see Sect. \ref{quantumcognition}).

The second insight came from the application of our quantum cognition
approach to the decision making process occurring in an Ellsberg-type
situation. In such a decision process, there is a contextual influence of a
cognitive, not physical, nature having its origin in the way the mind of the
person involved in the decision, i.e. a choice between two acts, relates to
the situation that is the subject of the decision making, i.e. the Ellsberg
situation. This led us to represent both the Ellsberg and Machina paradox
situations by using the mathematical formalism of quantum mechanics in
Hilbert space \cite{ast2014,hk2009,ast12}.

Let us firstly introduce what we call the `cognitive Ellsberg entity',
namely, the situation of an urn containing 30 red balls and 60 black and
yellow balls, the latter in unknown proportion. We assume that this entity
is in a specific state $p_{v}$, represented by a unit vector $|v\rangle $ of
the Hilbert space ${\mathbb{C}}^{3}$  on the field of complex numbers $\mathbb{C}$.\footnote{The choice of a 3-dimensional space is justified by the fact that three elementary and independent events appear in this Ellsberg urn.} Let us
then consider a `color measurement' on the Ellsberg entity, with three
possible outcomes, `red', `yellow' and `black', which we describe by the
context $g$ and represent by the self-adjoint operator $\mathcal{G}$. The
corresponding probability of getting one of these three outcomes in a
measurement of $g$ in the state $p_{v}$ can be interpreted as a subjective
probability, but such a probability is calculated by using standard rules
for quantum probability. Of course, the probability of getting `red' must be
1/3, if we want to represent the canonical three-color Ellsberg example. We realize at
once that, already at this stage, the presence of ambiguity is formalized by
means of a quantum state and a quantum probability measure for the events
occurring in this state. Moreover, this state can change under the influence of a
specific cognitive context, which reflects the different preference of
agents towards ambiguity (this perspective is getting more and more accepted in the scientific community, including mainstream psychologists; please see \cite{pnas}).
Let us then come to the acts $f_{1}$, $f_{2}$, $f_{3}$ and $f_{4}$. They are
represented by the self-adjoint operators ${\mathcal{F}}_{1}$, ${\mathcal{F}}%
_{2}$, ${\mathcal{F}}_{3}$ and ${\mathcal{F}}_{4}$, respectively, on ${%
\mathbb{C}}^{3}$. 
We can calculate the expected utility associated with each act $%
f_{i}$, $i=1,\ldots ,4$, in terms of the expected values of the operators ${%
\mathcal{F}}_{i}$ in the Ellsberg state $p_{v}$, again by following standard
quantum probabilistic rules. One can then show that, while the expected
utility associated with the acts $f_{1}$ and $f_{4}$ is independent of the
Ellsberg state, that is, this expected utility is `ambiguity-free', the
expected utility associated with the acts $f_{2}$ and $f_{3}$ depends on
this Ellsberg state, hence on subjective preferences towards ambiguity. This
means that it is possible to find suitable `Ellsberg superposition states',
that is, superpositions of two states incorporating different attitudes
towards ambiguity, such that the expected utility of $f_{1}$ is greater
(less) than the expected utility of $f_{2}$ and the expected utility of $%
f_{4}$ is greater (less) than the expected utility of $f_{3}$. This is in
perfect agreement with the typical pattern of response of individual agents
in Ellsberg-type paradoxes \cite{ast2014,hk2009,ast12}. One recognizes here a major novelty of our approach. The subjective
probabilities are not calculated through a fixed mathematical rule assigning
to individual events the same probability in each act. The subjective
probabilities instead `change' depending on the state of the cognitive Ellsberg entity, which incorporates subjective attitutes towards ambiguity.

The construction above shows that the Ellsberg situation can be represented
in a quantum-mechanical probability framework. But, to have an explicit
representation we needed to perform a real experiment on human participants.
This is exactly what we did, reporting its results in \cite{ast2014}. To
perform the experiment we sent out a questionnaire to several people,
including friends, relatives and students, to avoid statistical selection
biases. An extract of the text is as follows.

\emph{\ldots Imagine an urn containing 90 balls of three different colors:
red balls, black balls and yellow balls. We know that the number of red
balls is 30 and that the sum of the the black balls and the yellow balls is
60. The questions of our investigation are about the situation where
somebody randomly takes one ball from the urn.}

(i) \emph{The first question is about a choice to be made between two bets:
bet $f_1$ and bet $f_2$. Bet $f_1$ involves winning `10 euros when the ball
is red' and `zero euros when it is black or yellow'. Bet $f_2$ involves
winning `10 euros when the ball is black' and `zero euros when it is red or
yellow'. The question we would ask you to answer is: Which of the two bets,
bet $f_1$ or bet $f_2$, would you prefer?}

(ii) \emph{The second question is again about a choice between two different
bets, bet $f_3$ and bet $f_4$. Bet $f_3$ involves winning `10 euros when the
ball is red or yellow' and `zero euros when the ball is black'. Bet $f_4$
involves winning `10 euros when the ball is black or yellow' and `zero euros
when the ball is red'. The second question therefore is: Which of the two
bets, bet $f_3$ or bet $f_4$, would you prefer? \ldots}

Let us now analyze the obtained results. Our test on the Ellsberg paradox
problem involved 59 participants.\footnote{%
This is the typical number of participants in experiments on psychological
effects of the type investigated by Kahneman and Tversky, such as the
conjunction fallacy, and the disjunction effect \cite{tk1983,ts1992}. We
think that ambiguity aversion shares many features with these psychological
effects, as discussed in Sect. \ref{quantumcognition}.} The answers of the
participants were distributed as follows: (a) 34 participants preferred acts $%
f_1 $ and $f_4$; (b) 12 participants preferred acts $f_2$ and $f_3$; (c) 7
participants preferred acts $f_2$ and $f_4$; (d) 6 participants preferred acts $f_1$
and $f_3$. This makes the weights with preference of act $f_1$ over act $f_2$
to be 0.68 against 0.32, and the weights with preference of act $f_4$ over
act $f_3 $ to be 0.69 against 0.31. Hence, 46 participants over 59, that is,
78\%, chose the combination of act $f_1$ and act $f_4$ or act $f_2$ and act $%
f_3$. This inversion of preferences cannot be explained by SEUT, and our
finding replicated the most commonly observed choice pattern in the
three-color urn.

Participants' choices above can be represented in a quantum-mechanical
framework. Indeed, let us firstly consider the choice to bet on $f_{1}$ or
on $f_{2}$. This is a choice with two possible outcomes, say $o_{1}$ and $%
o_{2}$, hence it can be described as a measurement giving $o_{i}$ if act $%
f_{i}$ is chosen, $i=1,2$. The simplest outcomes are $o_{1}=+1$ and $o_{2}=-1
$. We represent the measurement associated with the first bet by the
self-adjoint operator ${\mathcal{O}}_{12}$. Let us then consider the choice
to bet on $f_{3}$ or on $f_{4}$. This is a choice with two possible outcomes
too, say $o_{3}$ and $o_{4}$, hence it can be described as a measurement
giving $o_{i}$ if act $f_{i}$ is chosen, $i=3,4$. The simplest outcomes are $%
o_{3}=+1$ and $o_{4}=-1$. We represent the measurement associated with the
second bet by the self-adjoint operator ${\mathcal{O}}_{34}$. We proved in 
\cite{ast2014} that the operatorial relation ${\mathcal{O}}_{12}\cdot {%
\mathcal{O}}_{34}={\mathcal{O}}_{34}\cdot {\mathcal{O}}_{12}$ holds. This
means that the corresponding `choice measurements' are compatible, or
commutative, in the sense that they can be measured together, and no order
effects should appear if we reverse the order of questions in our experiment
(see Sect. \ref{quantumcognition}). But, we also proved that the possibility
of representing our experimental data by compatible measurements for the
bets relies crucially on our choice of the Hilbert space ${\mathbb{C}}^{3}$
over complex numbers ${\mathbb{C}}$ as a modeling space. Indeed, if a
Hilbert space over real numbers is attempted, no compatible measurements for
the bets and the data can be constructed any longer \cite{ast2014}. 

The above result is relevant, in our opinion, since it shows that quantum
structures can be validly invoked in the Ellsberg paradox. Indeed, the
existence of compatible measurements to represent decision makers' choices
among the different acts in our experiment is a direct consequence of the
fact that we used a complex Hilbert space as a modeling space. 
If one instead uses a real Hilbert space, then the collected experimental data
cannot be reproduced by compatible measurements. Hence, two alternatives are
possible. Either one requires that compatible measurements occur in an
Ellsberg-type situation, and then one has to accept a complex Hilbert space
representation where ambiguity is incorporated into superposed quantum
states, and these superpositions are of the `complex type'. Hence entailing
genuine interference since whenever a superposed state vector is constructed
with complex (non-real) coefficients, the quantum effect of interference is
always at play. Alternatively, one can use a representation in a real
Hilbert space but, then, one should accept that an Ellserg-type situation
cannot be reproduced by compatible measurements. In either case, the
appearance of quantum structures, i.e. interference due to the presence of
genuine complex numbers, or incompatibility due to the impossibility to represent the data by compatible
measurements, seems unavoidable in the Ellsberg paradox situation.

Let us briefly summarize the novelties and possible advantages of using a
quantum-theoretic approach to model economic agents' decisions under
uncertainty. By introducing a more general quantum probabilistic framework
we are able both to reproduce the typical pattern that is observed in an
Ellsberg paradox situation, and to model a real decision making experiment
on this paradoxical situation. The representation of the decision maker's
subjective beliefs and probabilities by means of respectively a quantum
state vector and a quantum probability measure naturally capture the
presence of ambiguity in this type of situations. And, more important,
ambiguity aversion is accounted for by describing the decision maker's
choice as the result of a contextual interaction with the cognitive context.
In this respect, ambiguity aversion is one of the cognitive landscapes
driving human decisions under uncertainty, but other cognitive contexts may
well be present, and those can be modeled in our quantum framework. The
latter remark is important, and we think it deserves further explanation.

Firstly, our quantum-theoretic approach works well in the traditional
Ellsberg experiment (`single three-color urn'), but it perfectly reproduces
the other experiments considered by Ellsberg (`single four-color urn',
`double two-color urn').

Secondly, our approach is general enough to cope with different empirically
based human attitudes towards ambiguity. Indeed, many experiments were
performed after Ellsberg and, while most of them confirmed ambiguity
aversion \cite{ellsbergexp1}, some experiments could be explained in terms
of `ambiguity neutrality' and even `ambiguity attraction' \cite{ellsbergexp2}%
. This seems to confirm our insight that other cognitive contexts may play a
role in driving human decision making in these situations.

It is worth mentioning, to conclude, that our quantum-theoretic modeling is also
sufficiently general to cope with various generalizations of the Ellsberg
paradox, which are problematic, such as the Machina paradox and other
similar `ambiguity laden' situations \cite{machina2009,ast12,bdhp2011}. In
this respect, the preferences in the Machina reflection example in Sect. \ref{ellsberg} can be described by assuming that the 
cognitive context, mainly driven by `informational symmetry', determines a change in the state of the `cognitive Machina entity'. We have recently worked out a quantum-theoretic model of the Machina paradox situation where the state of the cognitive Machina entity is represented by a unit vector of the Hilbert space ${\mathbb C}^{4}$ -- here we have four elementary events -- and is bijectively associated with a subjective probability distribution. In other words, also in the Machina paradox situation, the subjective probability changes with the state and is influenced by the cognitive context. Our quantum-theoretic model in ${\mathbb C}^{4}$ reproduces informational symmetry and perfectly agrees with the data collected in \cite{aertssozzoMachina2015}. 

We are currently investigating the possibility of  a contextual
generalization of EUT and SEUT based on a quantum probability framework, which would provide a normative status to our approach.

\end{document}